\documentclass[12pt]{article}

\usepackage[dvips]{color}

\newcommand{\Black}{\color [rgb]{0,0,0}}

\newcommand{\Brown}{\color [rgb]{0.4,0.1,0.1}}

\def\be{\begin{equation}}
\def\ee{\end{equation}}

\def\TL{\hfil$\displaystyle{##}$}
\def\TR{$\displaystyle{{}##}$\hfil}
\def\TC{\hfil$\displaystyle{##}$\hfil}
\def\TT{\hbox{##}}
\def\seqalign#1#2{\vcenter{\openup1\jot
  \halign{\strut #1\cr #2 \cr}}}

\def\fixit#1{}

\def\mop#1{\mathop{\rm #1}\nolimits}
\def\tr{\mop{tr}}

\def\Vol{\mop{Vol}}
\def\Re{\mop{Re}}

\def\href#1#2{#2}  

\def\eqalign#1{\vcenter{\openup1\jot
    \halign{\strut\span\TL & \span\TR\cr #1 \cr
   }}}
\def\lbldef#1#2{\expandafter\gdef\csname #1\endcsname {#2}}
\def\eqn#1#2{\lbldef{#1}{(\ref{#1})}%
\begin{equation} \eqalign{#2} \label{#1} \end{equation}}
\def\eno#1{(\ref{#1})}
\begin{document}
\baselineskip=16pt
\pagestyle{plain}
\setcounter{page}{1}
\begin{titlepage}

\begin{flushright}
PUPT-2069 \\
hep-th/0212138
\end{flushright}
\vfil

\begin{center}
{\huge A universal result on central charges}
\vskip0.5cm
{\huge in the presence of double-trace deformations}
\end{center}

\vfil
\begin{center}
{\large Steven S. Gubser and Igor R. Klebanov}
\end{center}

$$\seqalign{\span\TL & \span\TT}{
& Joseph Henry Laboratories, Princeton University, Princeton, NJ 08544
}$$
\vfil

\begin{center}
{\large Abstract}
\end{center}

\noindent

We study large $N$ conformal field theories perturbed by relevant
double-trace deformations.  Using the auxiliary field trick, or
Hubbard-Stratonovich transformation, we show that in the infrared the
theory flows to another CFT.  The generating functionals of planar
correlators in the ultraviolet and infrared CFT's are shown to be
related by a Legendre transform.  Our main result is a universal
expression for the difference of the scale anomalies between the
ultraviolet and infrared fixed points, which is of order $1$ in the
large $N$ expansion.  Our computations are entirely field theoretic,
and the results are shown to agree with predictions from AdS/CFT.  We
also remark that a certain two-point function can be computed for all
energy scales on both sides of the duality, with full agreement
between the two and no scheme dependence.

\vfil
\begin{flushleft}
December 2002
\end{flushleft}
\end{titlepage}
\newpage
\renewcommand{\thefootnote}{\arabic{footnote}}
\setcounter{footnote}{0}
\tableofcontents

\section{Introduction}

The AdS/CFT correspondence is a duality between string or M-theory
backgrounds of the form $AdS_{d+1}\times X$ and conformal field 
theories in $d$ dimensions \cite{Malda,gkPol,WittenAdS}. 
There exist many checks of the duality based
on studying explicit models; the most extensively studied one is
the ${\cal N}=4$ SYM theory dual to type IIB string theory in
$AdS_5 \times S^5$ (for reviews, see \cite{MAGOO,Kleb,DHokerDan}). 
One may be bold enough to
argue that, whenever there exists an $AdS_{d+1} \times X$ solution
of string or M-theory, then it serves as a
constructive definition
of a $d$-dimensional
CFT, even if its conventional field theoretic formulation is
lacking.\footnote{This might even be argued for an $AdS_{d+1}$
background of some as-yet-unknown theory of quantum gravity with a
semi-classical limit in which Einstein gravity is recovered.}
The universal part of this construction is the $AdS_{d+1}$
space, while the ``details'' of the CFT, such as the number of 
supersymmetries, the global symmetries, etc., are encoded in
the compact space $X$. It is particularly interesting to
look for model-independent checks of the AdS/CFT duality which 
do not depend on $X$ explicitly. Examples of such checks
include studies of finite temperature theories via black holes in
AdS \cite{Peet,Wittentherm}, 
calculations of quark -- antiquark potential
\cite{Maldaloop,Rey}, and, quite recently,
a study of dimensions of high-spin operators \cite{gkPolTwo}.

In this paper we present a new general check 
of the duality which concerns the change
of the central charge under a flow produced by a double-trace
deformation. Our check applies to all models which contain scalar
fields in $AdS_{d+1}$ with
\eqn{range}{
-{d^2\over 4}  < m^2 L^2 \leq - {d^2\over 4} + 1
\,,
}  
so that both $\Delta_+$ and $\Delta_-$ dimensions are admissible
for the dual operator ${\cal O}$ \cite{kwTwo} ($\Delta_\pm$
are the two roots of $\Delta (\Delta -d) = (m L)^2$, where $L$ is the
radius of $AdS_{d+1}$).
Multi-trace deformations were first studied in the context of
AdS/CFT duality in \cite{ABSMulti}. The modification of the
AdS boundary conditions by such operators was presented in
\cite{WittenMulti,BSS} and further elaborated in
\cite{Mueck,Minces,Shomer}. The boundary conditions were used in 
\cite{WittenMulti} to argue that, whenever a relevant 
double-trace deformation ${\cal O}^2$ is added to the action, then
the theory flows from a
UV fixed point where ${\cal O}$ has dimension $\Delta_- < d/2$
to an IR fixed point where ${\cal O}$ has dimension 
$\Delta_+= d-\Delta_-$.\footnote{This type of flow is well-known
in $O(N)$ models in $2<d<4$ with fields $\phi^a$, $a=1,\ldots, N$
transforming in the fundamental representation \cite{WilsonKogut}. 
An $AdS$ dual of the flow produced by ``double-trace'' operator
$ (\phi^a \phi^a)^2 $
was recently proposed in terms of a higher-spin gauge theory 
in \cite{KP}. A $d=4$ model where this flow may take place is the
$SU(N)\times SU(N)$ superconformal gauge theory of \cite{kwOne} which
contains relevant double-trace operators.}
The difference in the
Weyl anomaly produced by this flow 
in even dimensions $d$ was calculated in \cite{GM}
using AdS methods; the result is a remarkably simple
universal formula which depends only on $\Delta_-$ and $d$.
In this paper we rederive this formula using field theoretic
methods. Our derivation serves as a new interesting
check of the duality, and it also sheds light on the origin of the
universality of the result of \cite{GM}.  

Indeed, the field theory computation doesn't even depend on the
existence of an anti-de Sitter dual---nor on supersymmetry, nor on
gauge symmetry.  It is a general result, for any even dimension $d$,
for any UV dimension $\Delta_- \in (d/2-1,d/2)$ of ${\cal O}$, and
depending only on having some form of large $N$ expansion.  Thus it
has some interest in its own right, apart from its value as a check of
AdS/CFT.
 
{}From the AdS point of view, the infrared CFT differs from the
ultraviolet CFT
only in one respect: the scalar field dual to ${\cal O}$ is quantized
with the conventional $\Delta_+$ boundary condition in the
IR, but with the unconventional $\Delta_-$ boundary condition
in the UV. This 2-fold ambiguity in quantization of scalar
fields in $AdS_{d+1}$
with $m^2$ in the range (\ref{range}) was originally found
in \cite{BF}, while its relevance for the AdS/CFT correspondence was
elucidated in \cite{kwTwo}.
In particular, \cite{kwTwo} presented a prescription for calculating
correlation functions of operators with dimension $\Delta_-$.
At the leading order in $N$, 
this prescription asserts that the 
generating functional of correlation functions in the theory 
where ${\cal O}$ has dimension $\Delta_-$ is related 
by a Legendre transform to the
corresponding object in the theory where 
${\cal O}$ has dimension $\Delta_+$. 
In this paper we shed new light on this prescription by
combining it with the proposal of \cite{WittenMulti}.
The combined proposal then states
that, in the large $N$ limit,
the ultraviolet CFT is related to the infrared CFT
produced by a relevant ${\cal O}^2$ operator through a Legendre transform.
We are able to derive this result in the field theory, using the
Hubbard-Stratonovich auxiliary field.
As a further step, we compute the leading correction in
the $1/N$ expansion, which follows from the
1-loop diagram for the auxiliary field and scales as $N^0$.
The anomalous part of this determinant exactly 
reproduces the AdS result of \cite{GM}, in
confirmation of the AdS/CFT duality. 
For this approach to work it is not necessary
for the ultraviolet CFT to be supersymmetric; even if it is, the
relevant double-trace interaction breaks the supersymmetry. 

We also remark that our treatment of the double-trace operators
parallels a similar treatment given in \cite{Aki} for $c\leq 1$
matrix models of 2-d quantum gravity deformed by operators
${\cal O}^2$. In particular, the
integral over the Hubbard-Stratonovich auxiliary variable was used in
\cite{Aki} to compare the $O(N^0)$ (torus) corrections to the free
energy in theories with two different scaling
dimensions of operator ${\cal O}$.

\section{Outline of the field theory calculation}
\label{Outline}

Consider a conformal field theory with a gauge-singlet
single-trace scalar
operator ${\cal O}$ whose dimension $\Delta$ falls in the range
$(d/2-1,d/2)$.  The lower limit of this range is the dimension of a
free scalar field, and it can be shown that a lower dimension for
${\cal O}$ would be inconsistent with unitarity.  Let us also assume
that in the undeformed conformal field theory, $\langle {\cal O}(x)
{\cal O}(0) \rangle = 1/x^{2\Delta}$ on ${\cal R}^d$, and that higher
point functions of ${\cal O}$ are suppressed by some sort of $1/N$
factors, where we may take $N$ large.  Then there is a general
argument that, in the large $N$ limit, the renormalization group flow
triggered by the relevant deformation ${f \over 2} {\cal O}^2$
terminates at an infrared fixed point where ${\cal O}$ is again a
scaling operator, but of dimension $d-\Delta$.  The argument proceeds
via a Hubbard-Stratonovich transformation, as follows.  Consider the
partition function
 \eqn{Zconsider}{
  Z_f[J] = \int {\cal D} \phi e^{-S[\phi] - 
   \int {f \over 2} {\cal O}^2 + \int J {\cal O}} = 
   \left\langle e^{-\int {f \over 2} {\cal O}^2 + \int J {\cal O}}
    \right\rangle_0
 }
where $f$ is a constant but $J$ may not be.  The notation $\int {\cal
D} \phi$ indicates path integration over all the degrees of freedom of
the conformal field theory, and $S[\phi]$ is the undeformed action.
The notation $\langle \cdots \rangle_0$ indicates an expectation value
in the undeformed conformal field theory (that is, with $J=f=0$).  The
Hubbard-Stratonovich transformation amounts to introducing an
auxiliary field $\sigma$:\footnote{The
determinant in \eno{AuxField} is a formal expression, defined so that
$\sqrt{\det\left( -{1 \over f} {\bf 1} \right)} \int {\cal D} \sigma
\, e^{{1 \over 2f} \sigma^2} = 1$.  The contour for $\sigma(x)$ should
be rotated to run along the imaginary axis to ensure convergence.  In
Lorentzian signature, the $\sigma$ field may be introduced in such a
way that the action remains real throughout, up to the usual
$i\epsilon$ terms.}
 \eqn{AuxField}{
  Z_f[J] = \sqrt{\det\left( -{\textstyle {1 \over f}} {\bf 1} \right)}
   \int {\cal D\sigma} 
    \left\langle e^{\int \left( {1 \over 2f} \sigma^2 + \sigma {\cal O} + 
     J {\cal O} \right)} \right\rangle_0 \,.
 }
The assumption that higher point functions of ${\cal O}$ are
suppressed now enters in a crucial way:
 \eqn{ExpApprox}{
  \left\langle e^{\int (\sigma+J) {\cal O}} \right\rangle_0 \approx
   e^{{1 \over 2} \left\langle 
   \left( \int (\sigma+J) {\cal O} \right)^2 \right\rangle_0} \,,
 }
up to some $1/N$ corrections.  The remaining $\sigma$ integral
required for computing $Z_f[J]$ is now strictly Gaussian.  If we now
define three linear operators by the relations
 \eqn{GKQ}{\seqalign{\span\TC}{
  (\hat{G}\sigma)(x) = \int d^d \xi \sqrt{g} 
   \langle {\cal O}(x) {\cal O}(\xi) \rangle_0 \sigma(\xi)  \cr
  \hat{K} = 1 + f\hat{G} \qquad 
   \hat{Q} = -{1 \over f} \left( \hat{K}^{-1} - 1 \right) = 
   {\hat{G} \over 1 + f\hat{G}} \,,
 }}
then it is straightforward to show that
 \eqn{ZJf}{
  Z_f[J] = {1 \over \sqrt{\det \hat{K}}} e^{{1 \over 2} 
   \langle J,\hat{Q} J \rangle} \,.
 }
Thus in particular, the two-point function for ${\cal O}$ in the
presence of the deformation is
 \eqn{TwoO}{
  \langle {\cal O}(x) {\cal O}(0) \rangle_f = 
   {\partial^2 \log Z_f[J] \over \partial J(x) \partial J(0)} = Q(x,0) \,,
 }
where $Q(x,0)$ is the position space representation of the operator
$\hat{Q}$. In the dual AdS treatment the same formula for the
two-point function was obtained in \cite{Mueck}.

The three linear operators, $\hat{G}$, $\hat{K}$, and $\hat{Q}$, are
diagonal in a momentum space basis for functions on ${\cal R}^d$.
Explicitly,
 \eqn{GKQpspace}{
  G(k) &= \int d^d x {e^{i k \cdot x} \over x^{2\Delta}} = 
   2^{d-2\Delta} \pi^{d/2} {\Gamma\left( {d \over 2} - \Delta \right)
    \over \Gamma(\Delta)} k^{2\Delta-d}  \cr
  K(k) &= 1 + f G(k) \qquad Q(k) = {G(k) \over 1 + f G(k)} \,.
 }
Expanding $Q(k)$ for small wave-numbers, we find
 \eqn{QExpand}{\seqalign{\span\TL & \span\TR & \span\TT}{
  Q(k) &= {1 \over f} - {1 \over f^2 G(k)} +
   {1 \over f^3 G(k)^2} - \ldots &\qquad\hbox{for $f G(k) \gg 1$} \cr
  \langle {\cal O}(x) {\cal O}(0) \rangle_f &\approx 
   -{1 \over f^2 \pi^d} {\Gamma(\Delta) \Gamma(d-\Delta) \over
    \Gamma\left({d \over 2} - \Delta\right) 
    \Gamma\left(\Delta - {d \over 2}\right)}
    {1 \over x^{2(d-\Delta)}} &\qquad
  \hbox{for $x \gg f^{-{1 \over d-2\Delta}}$.}
 }}
The position space expression shown comes from the $-1/f^2 G(k)$ term,
which is the leading non-analytic behavior of $Q(k)$ in the small $k$
limit.  Using the relation $\Gamma(x) \Gamma(1-x) = \pi/\sin\pi x$ and
the constraint $\Delta \in (d/2-1,d/2)$, it is easy to show that the
position space expression for $\langle {\cal O}(x) {\cal O}(0)
\rangle_f$ is positive.

The power law behavior for $\langle {\cal O} {\cal O} \rangle_f$ in the
infrared is {\it prima facia} evidence for an infrared fixed point.
Furthermore, we find that the dimension of operator ${\cal O}$
has changed from $\Delta$ in the UV to $d-\Delta$ in the IR,
in agreement with the reasoning presented in \cite{WittenMulti}
on the AdS side of the duality.
In fact,
$\langle {\cal O}(x) {\cal O}(0)
\rangle_f$ in the infrared limit and the original two-point function
$\langle {\cal O}(x) {\cal O}(0)\rangle_0$
are related by the Legendre transform prescription proposed in \cite{kwTwo}.
More generally, introduction of the 
Hubbard-Stratonovich auxiliary field explains why
the generating functionals 
of planar correlations functions in the infrared CFT, corresponding
to operator 
${\cal O}$ having
dimension $\Delta_+$, and in the UV
CFT, corresponding to dimension $\Delta_-$, are related by the Legendre
transform. To demonstrate this, define ${\cal W}[\sigma, h_i]$
to be the generating functional of correlators in the ultraviolet CFT: 
\eqn{genfun}{
e^{{\cal W}[\sigma, h_i]} = 
\left\langle e^{\int (\sigma {\cal O} + \sum_i h_i {\cal A}_i) }
    \right\rangle_0
}
for single-trace operators ${\cal A}_i$. Using (\ref{AuxField})
and shifting the auxiliary field, $\tilde \sigma= \sigma+ J$,
we find
\eqn{newAuxField}{
  Z_f[J,h_i] = \sqrt{\det\left( -{\textstyle {1 \over f}} {\bf 1} \right)}
   \int {\cal D}\tilde\sigma
  e^{{\cal W}[\tilde \sigma, h_i]+ \int  {1 \over 2f} (\tilde \sigma-J)^2 } 
\,. }
Now it is convenient to rescale $J=f \tilde J$ and send $f$
to $\infty$, which corresponds to taking the IR limit.
Discarding the term $f\tilde J^2/2$ which contributes only a contact term,
we find
\eqn{newnewAuxField}{
  Z_f[\tilde J,h_i] \sim
   \int {\cal D}\tilde\sigma
  e^{{\cal W}[\tilde \sigma, h_i] +\int \tilde \sigma \tilde J} 
 \,.}
This expression, which is analogous to the result of \cite{Aki},
may be used to generate the $1/N$ expansion in the infrared CFT.
To pick out the planar limit, it is sufficient to find the saddle point
for $\tilde \sigma$, so that the IR generating
functional is a Legendre transform of the UV one, 
\eqn{Leg}{
\log Z_f [\tilde J,h_i] = 
{\cal W}[\tilde \sigma, h_i] + \int \tilde \sigma \tilde J 
}
where
\eqn{newLeg}
{\tilde J= - {\delta {\cal W}[\tilde \sigma, h_i]\over \delta \tilde \sigma}
\,.
}
Conversely, the UV generating functional is a Legendre transform
of the IR one, in agreement with the results of 
\cite{kwTwo,WittenMulti}:
\eqn{invLeg}{
{\cal W}[\tilde \sigma, h_i] =
\log Z_f [\tilde J,h_i]- \int \tilde \sigma \tilde J 
\,,\qquad\qquad \tilde \sigma = 
{\delta \log Z_f [\tilde J,h_i] \over \delta \tilde J}
\,.}

Now, we go on to consider ${\cal O}(N^0)$
corrections to the leading order result
contained in the factor ${1 \over \sqrt{\det \hat{K}}}$
which comes from fluctuations of the auxiliary field 
away from its classical value.
In order to extract the central charge at the IR fixed point, it
will be convenient to consider the theory on a sphere $S_R^d$ of
radius $R$.  Conformally mapping ${\bf R}^d$ to $S_R^d$ (via a
stereographic projection, for example), one can easily show that
 \eqn{Sdtpf}{
  \langle {\cal O}(x_1) {\cal O}(x_2) \rangle = 
   {1 \over s(x,x')^{2\Delta}} \,,
 }
where $s(x,x')$ is the chordal distance between points $x$ and $x'$ on
$S^d$: that is, $s(x,x') = 2R \sin(\theta/2)$, where $\theta$ is the
angle between $x$ and $x'$.  The operators $\hat{G}$, $\hat{K}$, and
$\hat{Q}$ are now diagonal in a basis of spherical harmonics on
$S^d$.  An efficient way to find the eigenvalues $g_\ell$ of $\hat{G}$
is to expand
 \eqn{sExpand}{
  {1 \over s(x,x')^{2\Delta}} = 
   \sum_{\ell,m} g_\ell Y_{\ell m}^*(x) Y_{\ell m}(x') \,,
 }
where $\ell$ is the principal angular quantum number and $m$
collectively denotes all the magnetic quantum numbers.  The $Y_{\ell
m}(x)$ are assumed to include a factor of $R^{-d/2}$, so that when
they are squared and integrated over $S^d_R$, the result is unity.
For any $x'$, we have
 \eqn{glvalues}{
  g_\ell = {1 \over Y_{\ell m}(x')} 
   \int d^d x \sqrt{g} \, {1 \over s(x,x')^{2\Delta}} 
   Y_{\ell m}(x) \,.
 }
The eigenvalue $g_\ell$ has no $m$ dependence because of $SO(d+1)$
invariance.  Now we may choose $x'$ to be the north pole of $S^d$,
$\theta=0$.  Using the fact that on $S^d$, $Y_{\ell 0}(\theta)$ is
proportional to the Gegenbauer polynomial
$C_\ell^{(d-1)/2}(\cos\theta)$, we have
 \eqn{glint}{
  {g_\ell \over R^{d-2\Delta}} &= {\Vol S^{d-1} \over C_\ell^{(d-1)/2}(1)}
   \int_{-1}^1 dz (1-z^2)^{(d-2)/2} (1-z)^{-\Delta} C_\ell^{(d-1)/2}(z)  \cr
   &= {\Vol S^{d-1} \over C_\ell^{(d-1)/2}(1)} (-1)^\ell 
    2^{d-1-\Delta} \Gamma(d/2) {(\ell+d-2)!\over \ell! (d-2)!} 
    {\Gamma(1-\Delta) \Gamma\left( {d \over 2}-\Delta\right)
      \over \Gamma(1-\ell-\Delta) \Gamma(d+\ell-\Delta)}  \cr
   &= (-1)^\ell \pi^{d/2} 2^{d-\Delta} 
    {\Gamma(1-\Delta) \Gamma\left( {d \over 2} - \Delta \right)
     \over \Gamma(1-\ell-\Delta) \Gamma(d+\ell-\Delta)}  \cr
   &= \pi^{d/2} 2^{d-\Delta} {\Gamma\left( {d \over 2}-\Delta \right )\over
     \Gamma(\Delta)} {\Gamma(\ell+\Delta)\over \Gamma(d+\ell-\Delta)} \,,
 }
where we have used 
 \eqn{cellValue}{
  C_\ell^{(d-1)/2}(1)= {(\ell+ d-2)!\over \ell! (d-2)!} \,,
 }
and
 \eqn{volSd}{
  \Vol S^{d-1} = {2\pi^{d/2}\over \Gamma(d/2)} \,.
 }

The convenience of working on $S^d$ stems from the fact that the
central charge may be defined through the integrated one-point
function of the trace of the stress tensor: for a conformal field
theory,
 \eqn{Tmumu}{
  \left\langle \int_{S^d_R} d^d x \sqrt{g} \, T^\mu_\mu 
   \right\rangle = c \,,
 }
where the radius $R$ is arbitrary.  The central charge so defined
differs by factors of order unity from the usual definitions (where,
for example, in two dimensions $c=1$ for a free boson, and in four
dimensions $c=N^2/4$ for ${\cal N}=4$ $U(N)$ gauge theory).  For a
flow between conformal fixed points, the one-point function in \Tmumu\
will depend on $R$, interpolating between $c_{UV}$ for small $R$ and
$c_{IR}$ for large $R$.  For CFT's in even dimensions greater than
$2$, there are several distinct central charges, corresponding to
different possible Lorentz-invariant counterterms of the appropriate
dimension.  In CFT's with anti-de Sitter duals, at large $N$ it
follows from the considerations of \cite{HennSken} that all these
central charges are related in a rigid way.  This need not hold at
subleading order in $N$.  Cardy has conjectured \cite{Cardy} an analog
in four dimensions of Zamolodchikov's c-theorem for the central charge
appearing in \Tmumu.  We believe that this is the identical central
charge that was computed by supergravity methods, up to one loop, in
\cite{GM}.  At any rate, what we wish to do in the next section is to
compute
 \eqn{TmumuAgain}{
  c_{IR}- c_{UV}
 = \left\langle \int_{S^d_R} d^d x \sqrt{g} \, T^\mu_\mu 
   \right\rangle_f 
 - \left\langle \int_{S^d_R} d^d x \sqrt{g} \, T^\mu_\mu 
   \right\rangle_0 
= {1 \over d} R {\partial \over \partial R}(W_f[R]- W_0[R])
 }
where $W_f[R]\equiv \log Z_f[J=0,S_R^d]$, and
$Z_f[J=0,S_R^d]$ is the partition function of the CFT on the
sphere $S_R^d$, deformed by $\int {f \over 2} {\cal O}^2$.

\section{Determinant calculation on $S^d$}

With the result
 \eqn{gellAgain}{
  g_\ell = R^{d-2\Delta} 
   \pi^{d/2} 2^{d-\Delta} {\Gamma\left( {d \over 2}-\Delta \right )\over
     \Gamma(\Delta)} {\Gamma(\ell+\Delta)\over \Gamma(d+\ell-\Delta)}
 }
in hand, we wish to compute
 \eqn{Wdef}{
  W_f[R]- W_0[R] & = -{1 \over 2} \tr\log \hat{K}  \cr
   &= -{1 \over 2} \sum_{\ell=0}^\infty M_d(\ell) \log k_\ell
    = -{1 \over 2} \sum_{\ell=0}^\infty M_d(\ell) \log (1 + f g_\ell) \,,
 }
where $M_d(\ell)$ is the degeneracy of states with angular momentum
$\ell$ on $S^d$:
 \eqn{multip}{
  M_d(\ell) = {(\ell + d-2)! (2\ell + d-1)\over \ell! (d-1)! } \,.
 }
(This is the dimension of the irreducible representation of $SO(d+1)$
formed as the symmetric traceless part of $\ell$ fundamental vector
representations).  In the limit $f R^{d-2\Delta} \to 0$ (a very small
sphere, or hardly any deformation), $W_f[R] \to
W_0[R] $.  In the opposite limit, $f R^{d-2\Delta} \to \infty$, where we
are probing the infrared properties of the theory, the central charge
can be read off from the coefficient of $\log R$ in an expansion of
$W_f[R]$.  The reason for this is that the derivative in \TmumuAgain\
picks out $\log R$.

Dropping an overall constant which is independent of both $R$ and
$\Delta$, we find for large $R$ that
 \eqn{Wexpand}{
  W_f[R]- W_0[R] = -{1 \over 2} \sum_{\ell=0}^\infty M_d(\ell) \log g_\ell \,.
 }
Because the factors $\pi^{d/2} 2^{d-\Delta} \Gamma\left( {d \over 2} -
\Delta \right) / \Gamma(\Delta)$ in $g_\ell$ do not depend on $\ell$
or $R$, they will not contribute to the scale anomaly \TmumuAgain.
(The reader can check this explicitly once we have introduced our
regulation scheme).  Thus we are left with the computation of
 \eqn{WVDef}{
  W_f[R] = -{1 \over 2} \sum_{\ell=0}^\infty M_d(\ell)
   \log\left( R^{d-2\Delta} {\Gamma(\ell+\Delta) \over 
   \Gamma(d+\ell-\Delta)} \right)
   = -{1 \over 2} (V_1+V_2) \,,
 }
where 
 \eqn{Vonetwo}{
  V_1 = (d-2\Delta) (\log R) \sum_{\ell=0}^\infty M_d(\ell) \,,\qquad\quad
  V_2 = \sum_{\ell=0}^\infty M_d(\ell) 
   \log {\Gamma(\ell+\Delta) \over \Gamma(\ell+d-\Delta)} \,.
 }
Naively, it seems that only $V_1$ has $\log R$ dependence.  This would
be in contradiction with the results of \cite{GM}, because it
would mean that the central charge would depend linearly on
$d-2\Delta$.  The problem is that, because $M_d(\ell)$ is a polynomial
of degree $d-1$, all the sums in \Wdef, \Wexpand, \WVDef, and
\Vonetwo\ are divergent, and some regulator is required.  A regulator
must refer to some energy scale $\Lambda$ which remains fixed as we
differentiate with respect to $R$ in calculating $\langle \int
T^\mu_\mu \rangle$.  This introduces additional $R$ dependence beyond
what is apparent in \Vonetwo.

For reasons to be explained, we will settle eventually on
zeta-function regularization.  However, to appreciate the point about
the regulator introducing $R$-dependence to a sum over $\ell$, consider
the regulated sum
 \eqn{RegSum}{
  \sum_{\ell=1}^\infty \ell^\alpha e^{-\epsilon\ell} = 
   {\rm Li}_{-\alpha}(e^{-\epsilon})
 }
where $\epsilon^{-1} = R\Lambda$ and $\alpha$ is real.  When
$\alpha=-1$, the right hand side diverges as $-\log\epsilon =
\log(R\Lambda)$ in the $\epsilon \to 0$ limit.  For $\alpha<-1$, the
sum converges in the $\epsilon \to 0$ limit, while for $\alpha>-1$,
the sum (defined now via analytic continuation) has, at most,
divergences which are integer powers of $1/\epsilon$.  Evidently, for
$\alpha=-1$, there is logarithmic dependence on $R$ which persists in
the limit that the cutoff $\Lambda$ is removed.

This can be compared directly to a zeta-function regulator, which can
be motivated formally by considering
 \eqn{RegSumZeta}{
  \sum_{\ell=1}^\infty \ell^\alpha {1 \over \ell^s} = 
   \zeta(s-\alpha)
 }
for real $\alpha$.  Of course, the sum is again defined via analytic
continuation when $\Re s < \alpha+1$.  When $\alpha=-1$, there is a
pole in the expression on the right hand side at $s=0$ with residue
$1$.  For other real $\alpha$, there is no pole at $s=0$.  Comparing
with the remarks following \RegSum, one can conclude that for any
polynomial sum, extracting the residue from a zeta-function
regularization gives the coefficient of $\log(R\Lambda)$.  A
convenient feature of the zeta-function approach is that
 \eqn{LogSumZeta}{
  \sum_{\ell=1}^\infty \ell^\alpha (\log\ell) {1 \over \ell^s} = 
  {d \over d\alpha} \sum_{\ell=1}^\infty \ell^\alpha {1 \over \ell^s} = 
   -\zeta'(s-\alpha) \,,
 }
which has no residue.  Presumably it may be shown that $\ell^\alpha
(\log\ell)$ terms can also be dropped with the simpler exponential
regulator of the previous paragraph, but the argument seems less
transparent.

The last manipulation we need in order to be able to compute the
anomaly in any dimension is to set $x=\Delta-d/2$ and expand
 \eqn{GammExpand}{
  \log {\Gamma(\ell+\Delta) \over \Gamma(\ell+d-\Delta)} = 
   \log\Gamma(\ell+d/2+x) - \log\Gamma(\ell+d/2-x) = 
   2 \sum_{n=1\atop n\ \rm odd}^{\infty} {x^n \over n!} 
    \psi^{(n-1)}(\ell+d/2)
 }
where
 \eqn{PsiExpand}{
  \psi(z) = {d \over dz} \log\Gamma(z) = 
   \log z -{1\over 2z} -\sum_{k=1}^\infty {B_{2k}\over 2k} z^{-2k}
    = \log z + \sum_{s=0}^\infty  \zeta (-s) z^{-s-1} \,.
 }
Let us proceed now to the calculation of the anomaly in various cases.

\subsection{Vanishing of the anomaly in odd dimensions $d$}
\label{OddVanishes}

Because of the absence of appropriate generally covariant
counter-terms for violations of scale invariance, it is expected that
$\langle T^\mu_\mu \rangle = 0$ for a conformal theory in odd
dimensions, regardless of which curved manifold the theory is defined
on.

It is convenient here to define $k= \ell+(d-1)/2$.  The multiplicity
can be expressed as
 \eqn{MdOdd}{
  M_d(k-(d-1)/2) = {2\over (d-1)!} \prod_{i=0}^{(d-3)/2} (k^2 - i^2) \,.
 }
Now we observe that
 \eqn{VoneVanishes}{
  V_1 = (d-2\Delta) \log R \sum_{k=1}^\infty M_d(k-(d-1)/2)
 }
vanishes because $\zeta (-2 n)=0$ for all positive integer $n$.  The
reason we were able to shift the sum from $\ell=0$ to $\infty$ to a
sum from $k=1$ to $\infty$ is that $M_d(k-(d-1)/2)$ vanishes for
$k=1,2,\ldots,(d-1)/2$.

To show that $V_2$ also contributes nothing to the anomaly, we use the
large $k$ expansion
 \eqn{psiExpandAgain}{
  \psi(k+1/2) = \log k + \sum_{r=1}^\infty b_r k^{-2r} \,,
 }
where $b_1= 1/24$, $b_2=-7/960$, etc.  Since both $M_d$ and
$\psi(k+1/2)$ contain only even powers of $k$, the sums
 \be 
  \sum_{k=1}^\infty M_d(k-(d-1)/2) \psi^{(2n)} (k+1/2)
 \ee
contain no logarithmic divergences. Therefore, our calculation of the
anomaly indeed gives a vanishing result in all odd dimensions $d$.

The calculation of \cite{GM} gave a definite non-zero result for
the change in the bulk cosmological constant for odd $d$, but it is
unclear how to translate this in a crisp way into field theory terms.
Certainly the scale anomaly vanishes in the calculation of
\cite{HennSken} for odd dimensions.  The result of \cite{GM} may
correspond to some non-divergent terms in field theory that we are not
computing here.

\subsection{Vanishing of terms linear in $d-2\Delta$ in even dimensions $d$}
\label{LinearVanishes}

A general feature predicted by \cite{GM} is that there is no term
linear in $d-2\Delta$ in the expansion of the difference $c_{IR}-c_{UV}$ in
powers of $d-2\Delta$.  This can readily be shown in field theory, as
follows.

It is convenient here to define $k=\ell+d/2$.  The multiplicity
$M_d(k-d/2)$ vanishes for $k=1,2,\ldots,d/2$, so we are free to shift
a sum from $\ell=0$ to $\infty$ to a sum from $k=1$ to $\infty$.  We
may write
 \eqn{MExpand}{
  M_d(k-d/2) = \sum_{r=1}^{d-1} a_r k^r
 }
for some set of coefficients $a_r$.  Now, using the zeta-function regulator,
 \eqn{ZetaVone}{
  V_1=  (d-2\Delta) \log R \sum_{k=1}^\infty M_d =
  (d-2\Delta) \log R \sum_{r=1}^{d-1} a_r \zeta (-r) \,.
 }
On the other hand, the term in $V_2$ linear in $x=\Delta-d/2$ is
 \eqn{VtwoLinear}{
  (2\Delta-d) \sum_{k=1}^\infty M_d(k-d/2) \psi(k) &= 
   (2\Delta-d) \sum_{k=1}^\infty \left( 
    \sum_{r=1}^{d-1} a_r k^r \sum_{s=0}^\infty
     \zeta(-s) k^{-s-1} \right)  \cr
   &= (2\Delta-d) \log(R\Lambda) \sum_{r=1}^{d-1} a_r \zeta(-r) \,.
 }
In the first equality we have used \PsiExpand\ and dropped the
logarithmic term because it does not survive zeta-function
regularization.  In the second equality we have simply picked out the
$1/k$ terms in the product of the two interior sums.  Evidently, the
$\log R$ dependence cancels between \MExpand\ and \VtwoLinear.

\subsection{The cases $d=2$, $4$, $6$, and $8$}
\label{SeveralD}

It remains to compare the coefficients of non-linear odd powers in
$x=\Delta-d/2$ between our field theory calculation and the
$AdS_{d+1}$ calculation of \cite{GM}, where one finds an anomaly
proportional to
 \eqn{gmresult}{
  \int_0^x \prod_{i=0}^{(d-2)/2} (x^2 - i^2) \,.
 }
In the field theory calculation, only $V_2$ contributes.  We present
the calculation for a few even values of $d$, using the same
definition, $k=\ell+d/2$, as in section~\ref{LinearVanishes}.

First we consider $d=2$, where $M_2(\ell) = 2\ell+1 = 2k-1$, so that
the necessary sum is
 \eqn{CubicSum}{
  {x^3 \over 3} \sum_{k=1}^\infty (2k-1) \psi'' (k) \ \longrightarrow\
   -\log (R\Lambda) {2 x^3 \over 3} \,,
 }
where the arrow indicates that we have isolated the logarithmic term
by zeta-function regularization.  All higher powers of $x$ produce
convergent sums and cannot contribute to the anomaly.  The result
proportional to $x^3$ agrees with \cite{GM} up to overall
coefficient.

Now let us consider $d=4$, where the result found by \cite{GM} is
a particular sum of $(\Delta-2)^3$ and $(\Delta-2)^5$ behaviors.  We
have
 \eqn{MFourExpress}{
  M_4(\ell) = {(\ell+1)(\ell+2)(3+2\ell)\over 6}=
  {k(2 k^2 - 3 k+1)\over 6} \,,
 }
where $k=\ell+2$.  So the term of order $x^3$ is
 \eqn{xCubed}{
  {x^3\over 3} \sum_{k=1}^\infty 
   {k(2 k^2 - 3 k+1)\over 6}
   \left ( -{1\over k^2}- {1\over k^3} -{1\over 2 k^4} + \ldots \right ) 
  \ \longrightarrow\ \log(R\Lambda) {x^3 \over 18} \,.
 }
The term of order $x^5$ is evaluated similarly, and the total
anomalous part of $V_2$ is
 \eqn{AnomVTwoFour}{
  {1 \over 6} \log(R\Lambda) \left( {x^3\over 3} - {x^5\over 5} \right) \,,
 }
which agrees with \gmresult, again up to an overall normalization.

Now let us check the cases $d=6$ and $d=8$.  Defining $k=\ell+3$, we
have
 \eqn{MSixExpand}{
  M_6(\ell) = {2 k^5 - 5 k^4 + 5 k^2 - 2 k\over 5!} \,.
 }
By computations similar to \CubicSum\ and \xCubed, we find that the
anomalous part of $V_2$ is
 \eqn{VTwoSixAnom}{
  -{1 \over 180} \log (R\Lambda)
   \left ({x^7\over 7} - x^5 + {4 x^3\over 3}\right ) \,,
 }
where $x=\Delta-3$.  In the $d=8$ case, defining $k=\ell+4$, we have
 \eqn{MEightExpand}{
  M_8(\ell) = {(2k-1)(k-3)(k-2)(k-1)k(k+1)(k+2)\over 7!} \,,
 }
and the anomalous part of $V_2$ is
 \eqn{VTwoEightAnom}{
  -{1 \over 10080} \log(R\Lambda)
  \left ( {x^9\over 9} - 2 x^7 + {49 x^5\over 5}  - 12 x^3 \right ) \,.
 }
The results \VTwoSixAnom\ and \VTwoEightAnom\ again agree with
\gmresult.

Although we have not given a general argument for the agreement of the
field theory results with the simple AdS/CFT result \gmresult\ of
\cite{GM}, the cases that we have checked provide fairly
convincing evidence that in all even $d$ there is full agreement.

\section{A remark on the two-point function}
\label{TwoPoint}

One of the striking features of the general treatment in
section~\ref{Outline} of the field theory computation is that the
two-point correlator can be computed not just at the endpoints of the
renormalization group flow, but also at intermediate energy scales: on
${\bf R}^d$, the key formulas (recapitulating \TwoO\ and \GKQpspace)
are
 \eqn{RecapTG}{
  Q(k) &\equiv \int d^d x \, e^{-ik\cdot x} 
   \langle {\cal O}(x) {\cal O}(0) \rangle_f = 
   {G(k) \over 1 + fG(k)}  \cr
  G(k) &= 2^{d-2\Delta} \pi^{d/2} 
   {\Gamma\left( {d \over 2} - \Delta \right) \over \Gamma(\Delta)}
   k^{2\Delta-d} \,,
 }
where the original deformation is $S \to S + \int {f \over 2} {\cal
O}^2$.

On the other hand, the AdS treatment of the renormalization group flow
caused by the same deformation, initiated in \cite{WittenMulti} and
continued in \cite{Mueck,GM}, has also the capacity to
yield information about intermediate energy scales.
Indeed, in \cite{Mueck} the AdS methods were used to derive
the formula for the two-point function identical to (\ref{RecapTG}).
Below we rederive this result in a somewhat different way.
In
\cite{GM} the two-point function for the scalar $\varphi$ dual to
${\cal O}$ is given for arbitrary points in the bulk.  Parametrizing
Euclidean $AdS_{d+1}$ as
 \eqn{EucAdS}{
  ds^2 = {1 \over x_0^2} (dx_0^2 + dx_1^2 + \ldots + dx_d^2) \,,
 }
the two point function for points $(x_0,\vec{x})$, $(y_0,\vec{y})$
such that $x_0<y_0$ is given in \cite{GM} as
 \eqn{GEgm}{
  G_E(x,y;\tilde{f}) &= -\int {d^d k\over (2 \pi)^d}
   {e^{-i \vec{k} \cdot (\vec{x} - \vec{y})} 
    (x_0 y_0)^{d \over 2} K_{\nu}(ky_0) \over 
    (1 + (2/k)^{2\nu} \tilde{f} \, \Gamma(1 + \nu)/\Gamma(1 - \nu)) 
      L^{d-1}}  \cr\noalign{\vskip2\jot}&\qquad{}\times
    \left[ I_{-\nu}(k x_0) + \tilde{f} \, {\Gamma (1+\nu) \over \Gamma(1-\nu)} 
      \left( {2 \over k} \right)^{2\nu} I_\nu(kx_0) \right] \,,
 }
where $\nu = d/2-\Delta$ is between $0$ and $1$.\footnote{Note that
this sign for $\nu$ is opposite the one that often shows up in the
literature, for example \cite{gkPol}.  This is because $\nu$ is being
defined in reference to the UV scaling dimension of ${\cal O}$, rather
than the IR scaling dimension.}  Expanding $\varphi$ near the boundary
as
 \eqn{varphiExpand}{
  \varphi \sim \alpha(\vec{x}) x_0^{d-\Delta} + 
   \beta(\vec{x}) x_0^\Delta \,,
 }
the normalization of $\tilde{f}$ is fixed by writing the boundary
condition of \cite{WittenMulti} as $\alpha = \tilde{f}\beta$.

A correct though somewhat heuristic method to obtain the AdS/CFT
prediction for the two-point function $Q(k)$, up to an overall
normalization, is to start with the $k^{\rm th}$ Fourier component of
$G_E$, set $x_0=y_0 = \epsilon$, and extract the coefficient of the
leading non-analytic term in $\epsilon^2$ as $\epsilon \to 0$.  The
result is supposed to be $Q(k)$, up to an overall factor that may be
$\Delta$-dependent.  Dropping such factors, one obtains
 \eqn{AdSPredictsQ}{
  Q(k) \sim {k^{-2\nu} \over 1 + (2/k)^{2\nu}
   \tilde{f} \, \Gamma(1+\nu)/\Gamma(1-\nu)} \,.
 }
Only the term $I_{-\nu}(kx_0)$ inside square brackets in \GEgm\
contributes to \AdSPredictsQ.  Evidently, \AdSPredictsQ\ is in
agreement with \RecapTG\ (up to the overall normalization which we
have not endeavored to compute) provided we identify
 \eqn{fftilde}{
  \tilde{f} = {2\pi^{d/2} \over d-2\Delta} 
   {\Gamma\left( \Delta - {d \over 2} + 1 \right) \over \Gamma(\Delta)}
   f \,.
 }
It may be possible to normalize $\tilde{f}$ in an independent manner
and make a consistency check with \fftilde.  We presume that due
care would allow us to reconcile the overall normalization as well; but
since this can be studied entirely at the UV fixed point, and has only
to do with the $\int J{\cal O}$ couplings that have been well-explored
in other literature, we will not address the issue here.

The main point that the results of 
\cite{Mueck} and of this section illustrate is that the
AdS/CFT and field theory computations yield information about the
entire RG flow, with no scheme dependence in physical answers.

\section*{Acknowledgments}

We thank C.~Callan, L.~Rastelli, and E.~Witten for useful discussions.
I.R.K.\ is also grateful to the Institute for Advanced Study for
hospitality.  The work of S.S.G.\ is supported in part by the
Department of Energy under Grant No.\ DE-FG02-91ER40671.  The work of
I.R.K.\ is supported in part by the National Science Foundation under
Grant No.\ PHY-9802484.  Any opinions, findings, and conclusions or
recommendations expressed in this material are those of the authors
and do not necessarily reflect the views of the National Science
Foundation.

\bibliography{deform}
\bibliographystyle{ssg}

\end{document}